\documentclass[apjl]{emulateapj}

\usepackage{graphicx}
\usepackage{ulem}
\usepackage{url}
\usepackage{threeparttable}
\usepackage{xcolor}
\usepackage{soul}
\usepackage{booktabs}
\usepackage[]{natbib}

\usepackage{amsmath}
\usepackage[colorlinks, linkcolor=red, anchorcolor=green, citecolor=blue]{hyperref}

\graphicspath{{}{figures/}}
\usepackage{txfonts}

\def\purple#1   {{\textcolor{purple}{#1}}\ }
\def\red#1    {\textcolor{red}{#1}}
\def\new#1    {{\bf #1 }}
\def\cut#1    {\sout{#1} }

\def\kms    {\ifmmode{{\rm \ts km\ts s}^{-1}}\else{\ts km\ts s$^{-1}$}\fi}
\def\kms    {km\,s$^{-1}$\,}

\def\arcsec {\hbox{$^{\prime\prime}$}}

\def\cmt   {cm$^{-3}$\,}

\def\Kkms{K\,\kms }
\def\Kkmspc{K\,\kms pc$^2$}

\def\,{\thinspace}

\def\Lsun{$L_\odot$}

\def\Tkin   {{$T_{\rm kin}$}\,}

\def\Tmb    {{$T_{\rm mb}$}\,}
\def\Tsys   {{$T_{\rm sys}$}\,}

\def\Tex    {{$T_{\rm ex}$}\,}
\def\Tex    {\ifmmode{{T}_{\rm ex}}\else{$T_{\rm ex}$}\fi}

\def\Tkin   {$T_{\rm kin}$}

\def\HI         {H\,{\sc i}\,}

\,

\def\LIR     {$L_{\rm IR}$}
\def\LCO     {$L_{\rm CO}$}

\def\Lgas    {$L'_{\rm gas}$}
\def\LHCNft  {$L'_{\rm HCN\, J=4-3}$}

\def\LHCNoz  {$L'_{\rm HCN\, J=1-0}$}

\def\LHCOPft  {$L'_{\rm HCO^+\, J=4-3}$}
\def\LHCOP    {$L'_{\rm HCO^+}$}

\def\LCSss    {$L'_{\rm CS\, J=7-6}$}

\def\nht        {$n_{\mathrm{H}_2}$}
\def\HCOP       {HCO$^{+}$}
\def\C2H        {C$_{2}$H}
\def\HC3N       {HC$_{3}$N}
\def\H2O        {H$_2$O}

\def\HCNoz      {HCN\,$J$=1$\rightarrow$0}
\def\HCNft        {HCN\,$J$=4$\rightarrow$3}
\def\HCNtt        {HCN\,$J$=3$\rightarrow$2}
\def\HCOPft     {HCO$^+\,J$=4$\rightarrow$3}
\def\CSss          {CS\,$J$=7$\rightarrow$6}
\def\CSff          {CS\,$J$=5$\rightarrow$4}

\def\lsol   {\ifmmode{L_{\odot}}\else{$L_{\odot}$}\fi}
\,
\,
\,

\def \ncrit {$n_{\rm crit}$}

\def\Tastar{{$T_{\rm A}^\star$}}

\def\to      {$\rightarrow$}

\slugcomment{To appear in ApJL, 2014}

\shorttitle{Dense gas tracers and star formation laws in star-forming galaxies}
\shortauthors{Zhang et al.}

\begin{document}

\title{Dense gas tracers and star formation laws in active galaxies: 
APEX survey of  \HCNft\, \HCOPft, and \CSss}



\author{ Zhi-Yu Zhang \altaffilmark{1,2,3}, Yu Gao \altaffilmark{1}, Christian
Henkel\altaffilmark{2,4}, Yinghe Zhao \altaffilmark{5,1},   Junzhi
Wang \altaffilmark{6}, Karl M. Menten\altaffilmark{2}, \and Rolf
G{\"u}sten\altaffilmark{2} }

\affil{$^1$ Purple Mountain Observatory/Key Lab for Radio Astronomy, 2 West Beijing Road, Nanjing 210008, China}
\email{zyzhang@pmo.ac.cn}

\affil{$^2$ Max-Planck-Institut f{\"u}r Radioastronomie, Auf dem H\"ugel 69, D-53121 Bonn, Germany}

\affil{$^3$  University of the Chinese Academy of Sciences, 19A Yuquan Road, P.O. Box 3908, Beijing 100039, China}

\affil{$^4$ Astron. Dept., King Abdulaziz University, P.O.  Box 80203, Jeddah, Saudi Arabia }

\affil{$^5$ Infrared Processing and Analysis Center, California Institute of Technology, MS 100-22, Pasadena, CA 91125, USA} 

\affil{$^6$ Shanghai Astronomical Observatory, 80 Nandan Road, Shanghai 200030, China}


\begin{abstract} 
        We report \CSss, \HCNft, and \HCOPft\ observations in 20 nearby
        star-forming galaxies with the Acatama Pathfinder EXperiment 12-m
        telescope. Combined with 4 CS, 4 HCN, and 3 HCO$^+$ detections
        in literature, we probe the empirical link between the luminosity of
        molecular gas (\Lgas) and that of infrared emission (\LIR), up to the
        highest gas densities (10$^6$ - 10$^8$\cmt) that have been probed so
        far. For nearby galaxies with large radii, we measure the IR
        luminosity within the submm beam-size (14 --18$''$) to match the
        molecular emission. We find linear slopes for \LCSss-\LIR\ and
        \LHCNft-\LIR, and a slightly super-linear slope for \LHCOPft-\LIR. The
        correlation of \LCSss-\LIR\ even extends over eight orders of
        luminosity magnitude down to Galactic dense cores, with a fit of
        log(\LIR)$=1.00(\pm 0.01) \times$log(\LCSss) $+4.03(\pm0.04)$. Such
        linear correlations appear to hold for all densities $>10^4$\cmt,
        and indicate that star formation rate is not related to free-fall time
        scale for dense molecular gas. 
\end{abstract}

\keywords{ galaxies: evolution --- galaxies: ISM --- infrared:  galaxies ---
ISM: molecules ---  radio lines: galaxies }

\section{INTRODUCTION}\label{intro}

The star formation process constantly turns gas into stars. The
Kennicutt-Schmidt (K-S) law \citep{Kennicutt98} globally correlates the
surface-densities of star formation rate ($\Sigma_{\rm SFR}$ traced by
H$\alpha$) with total gas mass ($\Sigma$gas traced by CO and \HI), with a slope
$\alpha \sim 1.4\, (\Sigma_{\rm SFR} \propto \Sigma_{\rm gas}^\alpha)$.
However, quite a range of deviations from the K-S law have been observed, and
no unique slope was found in the \LCO-\LIR\ correlations as well
\citep[e.g.,][]{Liu2012, gs04b}.

Recent studies on star formation indicate that stars, especially the massive
stars, are predominantly formed in the dense cores of giant molecular clouds
\citep[e.g.,][]{e08}. Dense gas directly represents molecular content involved
in forming stars \citep[e.g.,][]{Lada12}, traced by the rotational transitions
of high dipole moment molecules (e.g., HCN and \HCOP), because of their high
critical densities\footnote{All critical densities ($n_{\rm crit}$) in this
letter are calculated with $n_{\rm crit} = \sum\limits_{u>l} A_{\rm u l} /
\sum\limits_{\rm u \neq l} C_{\rm ul}(T_{\rm kin})$ at \Tkin=100 K assuming
optically thin emission. $A_{\rm ul}$ and $C_{\rm ul}$ denote the Einstein
coefficient for spontaneous emission and the collision rate, respectively. All
state-to-state cross sections and rate coefficients are from the LAMDA Web site
(\url{http://home.strw.leidenuniv.nl/\~moldata/})
\citep{schoier2005}.}(\ncrit).  \citet[][]{gs04b,gs04a} find a tight
linear correlation between the luminosities of IR emission (\LIR\, tracing the
SFR) and HCN $J$=1$\rightarrow$0 (\LHCNoz\ tracing $M_{\rm dense}$) in
galaxies. This correlation even extends to Galactic dense cores undergoing high
mass star-formation \citep[e.g.,][]{Wu05}.   \citet{Lada12} argue that such
linear correlation is a fundamental relation for SFR and dense gas, and the
molecular gas with densities above $10^4$\cmt\ should follow this linearity.
Linear correlations have also been found in other dense gas tracers with
similar or higher \ncrit\ (e.g., \HCOP $J$=1$\rightarrow$0, HNC $J$=3\to2,
\CSff), in both Galactic dense cores and galaxies
\citep[e.g.,][]{Schenck11,Ma12,Reiter2011, Wang2011}.

\begin{figure*}[btp]
\includegraphics[angle=-180,scale=1.0,bb=1 515 510 790 ]{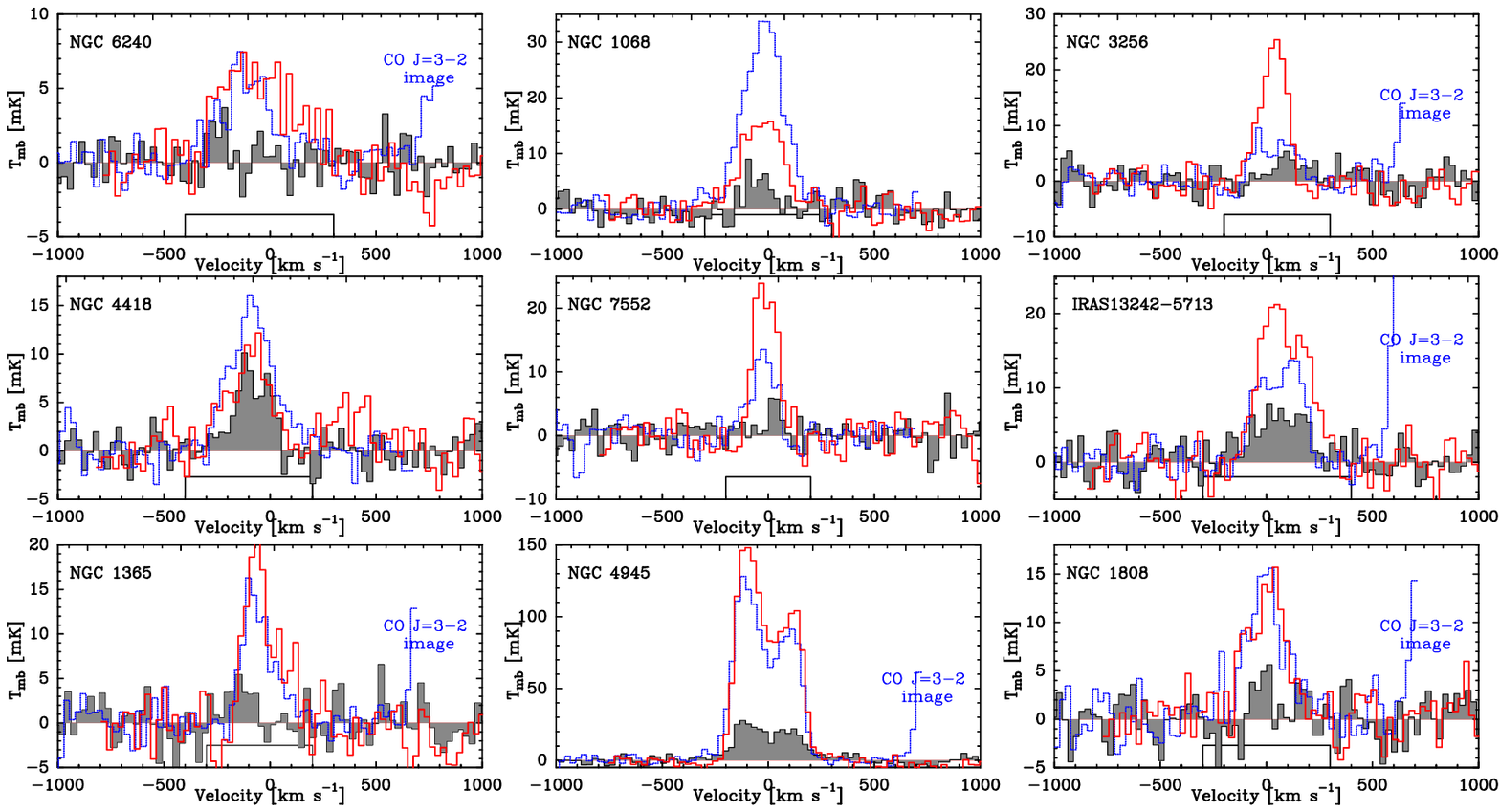}

\caption{Spectra of the strong detections of \HCOPft. Filled lines show \CSss,
red thick lines show \HCOPft, and blue dotted lines show \HCNft. All velocities
are labeled relatively to the velocity corresponding to their redshifts. }
\label{spectra1}                             
\end{figure*}

\begin{figure*}[btp]
\includegraphics[angle= 0,scale=1.0,bb=1 1 510 280]{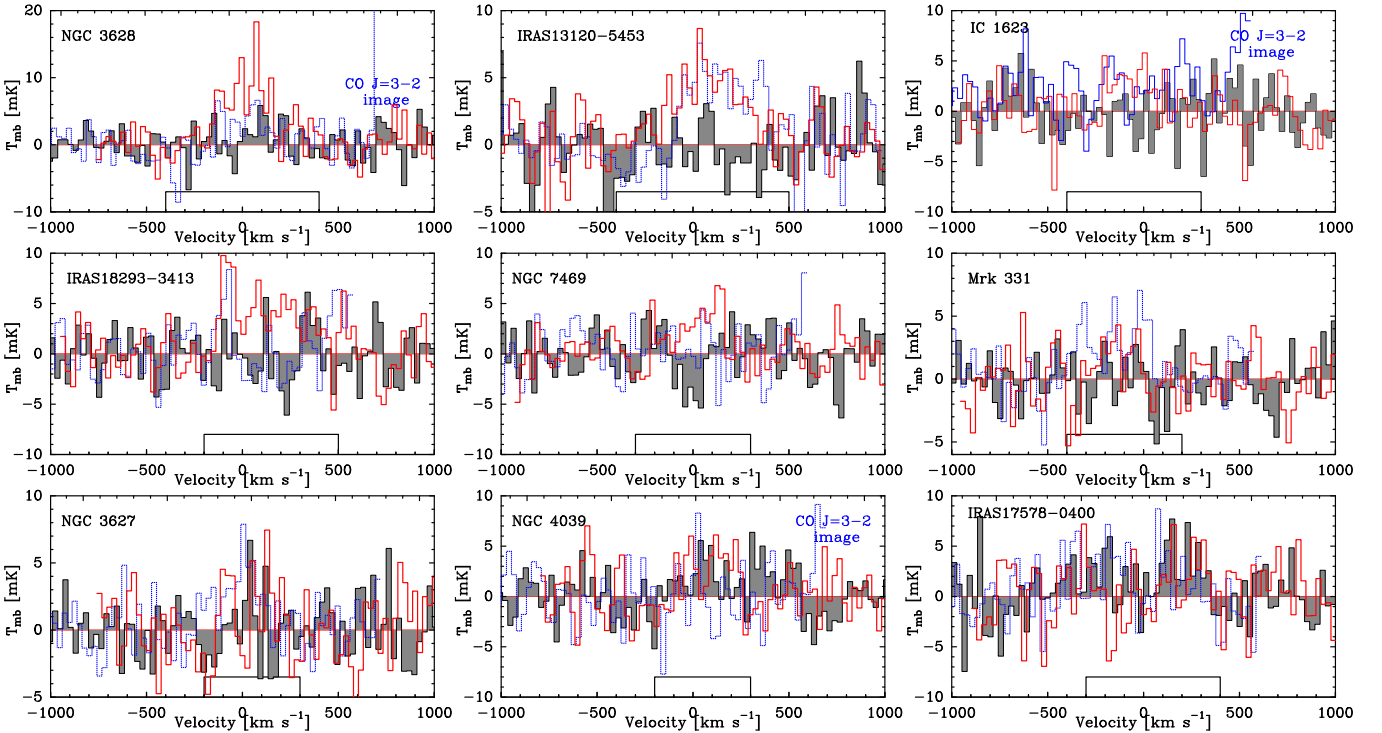}
\caption{Spectra of the weak detections of \HCOPft. The labels are the same as the Fig
\ref{spectra1}.}
\label{spectra2}
\end{figure*}

\begin{table*}
\caption{ Observed properties of the sample}\label{lineinfo}
\begin{center}
\scriptsize
\begin{tabular}{lcccccccccccc}\hline\hline
&Distance& \LIR\ & $I_{\rm CS(7-6)}$ &$I_{\rm HCN(4-3)}$ &$I_{\rm HCO^+(4-3)}$ & $L'_{\rm CS(7-6)}$& $L'_{\rm HCN(4-3)}$ & $L'_{\rm HCO^+(4-3)}$ & $R_{\rm SD}$&$C_{\rm aper}$&Band&Type$^g$ \\
 Source Name & (Mpc)& ($10^{10}$\Lsun)  & (\Kkms)  & (\Kkms)  & (\Kkms)  &  (10$^6$\Kkmspc)   &  (10$^6$\Kkmspc)    &  (10$^6$\Kkmspc)    &   & &($\mu$m) &   \\ 
\hline   
IC~1623         & 82.4 & 49.0  &    $<$  0.9 &         $<$  0.9 & 0.9$\pm$0.2 &    $<$  66   &     $<$  62     &    62$\pm$15   & 0.65 & 1.29 &100&SF\\ 
NGC~1068        & 14.4 & 20.4  & 1.0$\pm$0.2 &      5.0$\pm$0.2 & 2.7$\pm$0.2 & 2.5$\pm$0.5  & 10.7$\pm$0.4    &   5.8$\pm$0.4  & 0.16 & 1.16 &70 &AGN\\    
NGC~1365        & 18.3 & 10.5  & 0.6$\pm$0.2 &      1.7$\pm$0.2 & 2.9$\pm$0.2 & 2.2$\pm$0.7  &  5.9$\pm$0.7    &  10.0$\pm$0.7  & 0.25 & 1.16 &70 &AGN\\    
NGC~1808        & 12.3 & 6.92  &    $<$  0.6 &      2.5$\pm$0.2 & 2.2$\pm$0.2 &    $<$  1.0  &  3.9$\pm$0.3    &   3.4$\pm$0.3  & 0.46 & 1.29 &100&AGN\\    
NGC~2903        & 10.6 & 2.57  &    $<$  0.9 &         $<$  0.9 &    $<$  0.9 &    $<$  1.1  &     $<$  1.0    &      $<$  1.0  & 1.0  & --   &-- &SF\\    
NGC~3256        & 44.3 & 57.5  & 0.8$\pm$0.2 &      1.1$\pm$0.2 & 2.8$\pm$0.2 &  45$\pm$6    &   22$\pm$4      &    56$\pm$4    & 0.34 & 1.29 &100&SF\\ 
NGC~3627        & 11.0 & 2.88  &    $<$  0.7 &      0.8$\pm$0.2 & 0.7$\pm$0.2 &    $<$  1.0  &  1.0$\pm$0.3    &   0.9$\pm$0.3  & 0.08 & 1.29 &100&AGN\\    
NGC~3628        & 11.0 & 2.14  & 0.5$\pm$0.2 &      0.9$\pm$0.2 & 2.5$\pm$0.2 &    $<$  1.2  &  1.1$\pm$0.4    &   3.1$\pm$0.4  & 0.23 & --   &850&SF\\      
NGC~4039        & 25.0 & 9.33  &    $<$  1.2 &         $<$  0.7 & 0.9$\pm$0.3 &    $<$  8.2  &     $<$  4.5    &   5.8$\pm$1.8  & 0.04 & 1.29 &100 &SF\\    
NGC~4418        & 29.5 & 9.33  & 1.2$\pm$0.2 &      3.0$\pm$0.2 & 1.2$\pm$0.2 &11.5$\pm$2.0  & 26.8$\pm$2.0    &   8.9$\pm$1.8  & 1.0  & --   & --&AGN\\    
NGC~4945        & 3.8  & 2.82  & 7.3$\pm$0.3 &     23.4$\pm$0.3 & 9.9$\pm$0.3 & 1.2$\pm$0.05 &  3.6$\pm$0.05   &   4.5$\pm$0.05 & 0.40 & 1.29 &100&AGN\\    
IRAS13120-5453  & 128.3& 178   &    $<$  0.9 &      2.0$\pm$0.3 & 1.6$\pm$0.3 &    $<$  106  &  330$\pm$30     &   265$\pm$30   & 1.0  & --   &-- &SF\\ 
IRAS13242-5713  & 38.9 & 31.6  & 1.8$\pm$0.3 &      4.3$\pm$0.3 & 3.5$\pm$0.3 &  30$\pm$5    &   67$\pm$5      &    54$\pm$5    & 0.64 & 1.29 &100&SF\\  
NGC~6240        & 107  & 70.8  &    $<$  0.7 &      1.3$\pm$0.2 & 2.4$\pm$0.2 &    $<$  99   &  189$\pm$30     &   260$\pm$20   & 1.0  & --   &-- &SF\\
IRAS17578-0400  & 61.1 & 24.5  &    $<$  0.9 &      1.5$\pm$0.3 &    $<$  0.9 &    $<$  36.6 & 57.0$\pm$10     &      $<$  34   & 1.0  & --   &-- &SF\\ 
IRAS18293-3413  & 77.4 & 64.6  &    $<$  0.9 &      $<$ 0.9     & 1.6$\pm$0.3 &    $<$  58.6 &     $<$  55     &   116$\pm$20   & 1.0  & --   &-- &SF\\  
NGC~7469        & 71.6 & 46.8  &    $<$  0.9 &         $<$  0.9 & 1.0$\pm$0.2 &    $<$  33.5 &     $<$  30     &    37$\pm$10   & 1.0  & --   &-- &AGN\\ 
NGC~7552        & 19.5 & 8.91  & 0.6$\pm$0.2 &      1.3$\pm$0.2 & 2.4$\pm$0.2 & 1.3$\pm$0.4  &  5.1$\pm$0.4    &   9.4$\pm$0.4  & 0.59 & 1.29 &100&SF\\    
NGC~7771        & 63.0 & 26.9  &    $<$  0.9 &         $<$  0.9 &    $<$  0.9 &    $<$  40   &     $<$  36     &      $<$  36   & 0.55 & 1.29 &100 &SF\\ 
Mrk~331         & 80.9 & 33.9  &    $<$  0.9 &      1.1$\pm$0.2 &    $<$  0.9 &    $<$  43   &   60$\pm$13     &      $<$  40   & 1.0  & --   &-- &AGN\\ 
\hline                                                                                                                           
NGC~253$^a$     & 3.47 & 3.45  &     --      &         --       &     --      &0.35$\pm0.01$$^a$ & 8.9$\pm$1$^b$   & 7.7$\pm$1.0$^b$ & 0.18&1.33 &100 &SF\\
IC~342$^b$      & 3.93 & 1.08  &     --      &         --       &     --      &0.05$\pm0.02$$^a$ & 0.8$\pm$0.2$^c$ &      --         & 0.05&1.33 &100 &SF\\
Arp~220$^c$     & 79.9 & 162.2 &     --      &         --       &     --      &190 $\pm40$$^d$   & 950$\pm$190     & 170$\pm$40 $^d$ & 1.0 &--   &--  &SF\\
HE2-10$^d$      & 16.4  & 0.6  &      --     &         --       &     --      &0.21$\pm0.7$$^a$  &    --           &      --         & 0.29&1.33 &100 &SF\\
MRK~231$^e$     & 180.8 & 340  &     --      &         --       &     --      &    --            & 550$\pm$110$^e$ & 188$\pm$23$^f$  & 1.0 &--   &--  &AGN\\
\hline 
\end{tabular}\\
$^a$ \citep[][]{bayet09}. $^b$  \citep[][]{Knudsen07}. $^c$ \citep[][]{Jackson95}. $^d$  \citep[][]{gpg09}. $^e$ \citep[][]{PPP2007}. $^f$ \citep[][]{Wilson08}
$^g$ Galaxy types are found in NASA/IPAC Extragalactic Database (NED).  
\end{center}
\end{table*}

For gas tracers with \ncrit\ higher than HCN $J$=1$\rightarrow$0, the
slopes of the \Lgas-\LIR\ correlations are controversial. \citet{kt07} argue
that the mean densities in different types of galaxies and \ncrit\ of the
tracer change the slopes. Numerical simulations predict decreasing slopes
against increasing \ncrit, because of sub-thermal excitation conditions
\citep[i.e.,][]{ncs08,jnm09}. Observations of HCN, \HCOP, and CS
(typically, $J$$\le$3$\rightarrow$2) have been used to support these
and show sub-linear correlations \cite[e.g.,][]{bhl08, bns08,gc08}, where
\HCNtt\ has a slope of $\alpha\sim$ 0.8, following the prediction quite well
\citep{bns08}. Up to date, only few detections of higher-$J$
transitions of the above species have been reported
\citep[i.e.,][]{Knudsen07,Jackson95,gpg09,bayet09, Wilson08}. 

To better probe the densest molecular gas in galaxies, and to test the
predictions for higher \ncrit\ tracers, we therefore performed a survey of
\CSss, \HCNft, and \HCOPft\ in 20 nearby actively star-forming galaxies with
the Atacama Pathfinder EXperiment (APEX) 12-m telescope\footnote{This
publication is based on data acquired with the Atacama Pathfinder Experiment
(APEX). APEX is a collaboration between the Max-Planck-Institut f{\"u}r
Radioastronomie, the European Southern Observatory, and the Onsala Space
Observatory.}. With the advent of the {\it Herschel} space telescope,
we are able to obtain beam-matched IR luminosities in nearby galaxies, and to
compare them with data from single dish telescopes. In this letter, we
summarize our findings and compare them with the results from Galactic studies.
We adopt flat cosmology parameters ($\Omega_m$ =0.27, $\Omega_\Lambda$=0.73,
and $H_0$=71 \kms Mpc$^{-3}$; \citealp{sbd07}).

\section{SAMPLE, OBSERVATIONS and DATA REDUCTION}

We selected 20 galaxies from the  Infrared Astronomical Satellite
(IRAS) Revised Bright Galaxy Sample \citep{smk03}, with $S_{\nu}(100\, \mu m)
>$ 100 Jy, and declination $< 20 \degr$. In the analysis, we also include data
from literature \citep{Jackson95,Knudsen07,PPP2007,Wilson08,gpg09,bayet09},
which were mostly observed with the James Clerk Maxwell Telescope (JCMT) with a
beam-size (Full Width Half Power; FWHP) of 14$''$.  The sample encompasses
galaxies with \LIR\ from $10^{10}$\Lsun\ to $10^{12.5}$\Lsun, including nearby
normal galaxies, starbursts, and  Ultra Luminous Infrared Galaxies
(ULIRGs; \LIR$\ge10^{12}$\Lsun), with about half containing Active Galactic
Nuclei (AGNs).  Table \ref{lineinfo} lists the targets with integrated
intensity, distance, and luminosities. 

\begin{figure*}[!hbtp]
\includegraphics[angle=90,scale=0.4]{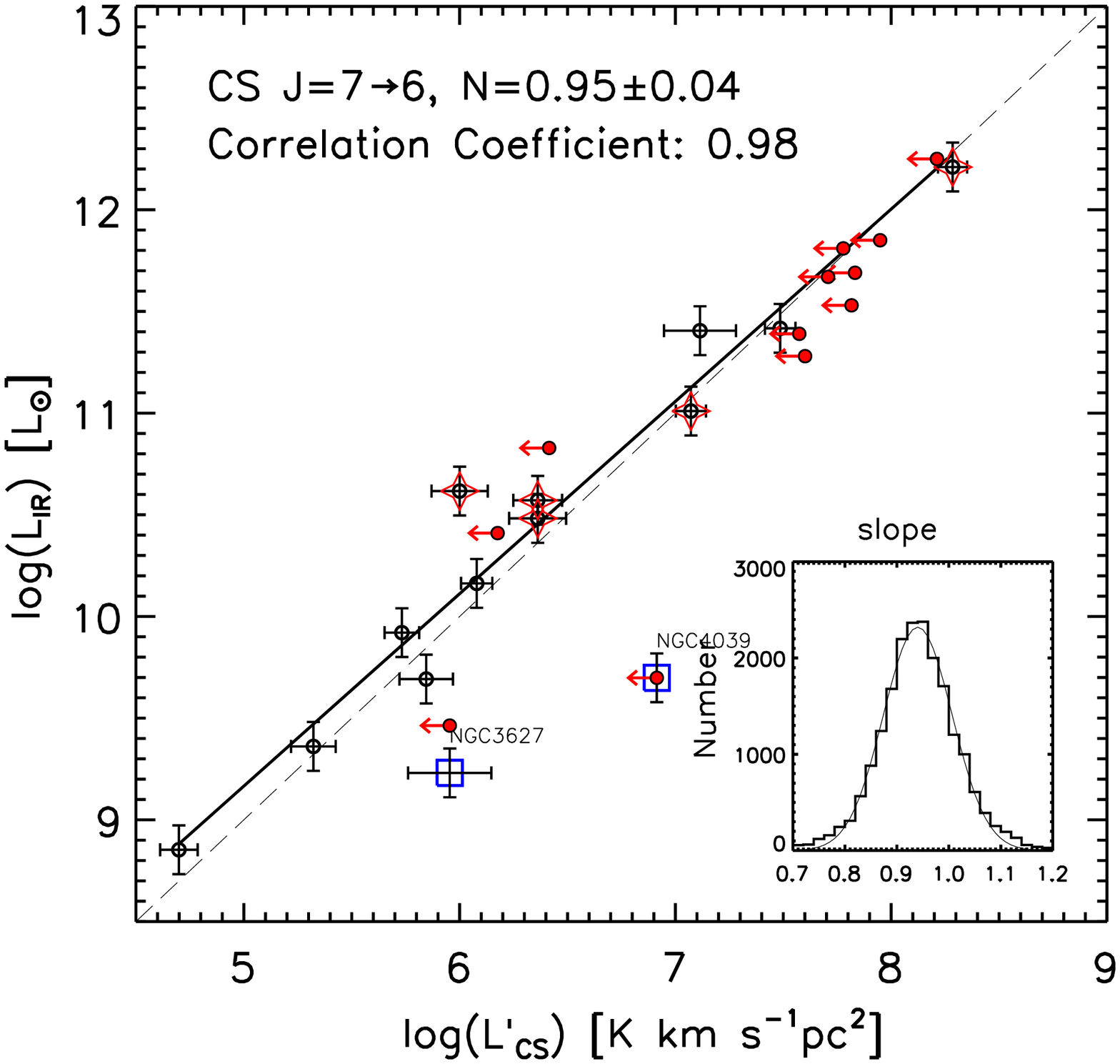} 
\includegraphics[angle=90,scale=0.4]{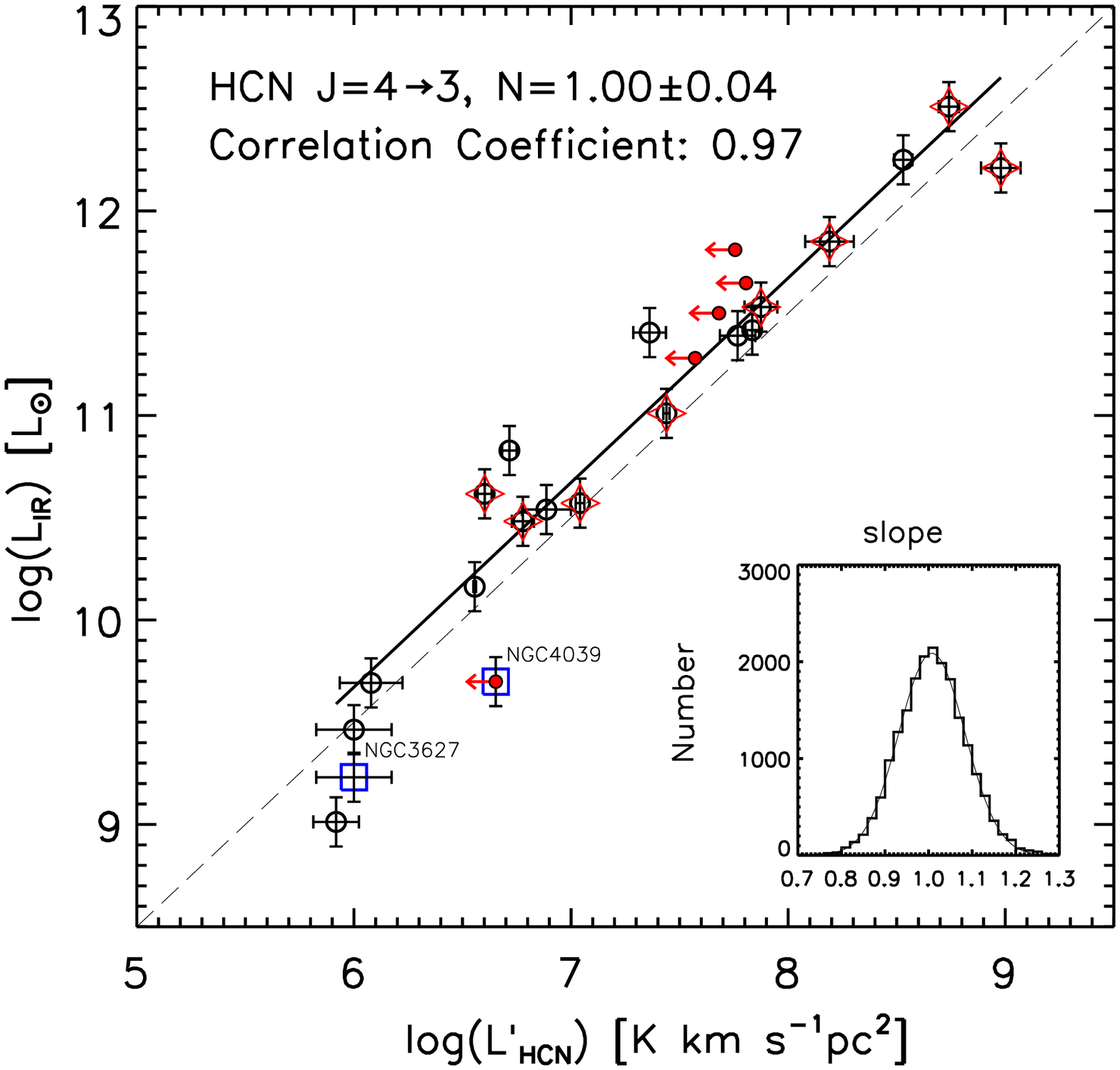} \\ 
\includegraphics[angle=90,scale=0.4]{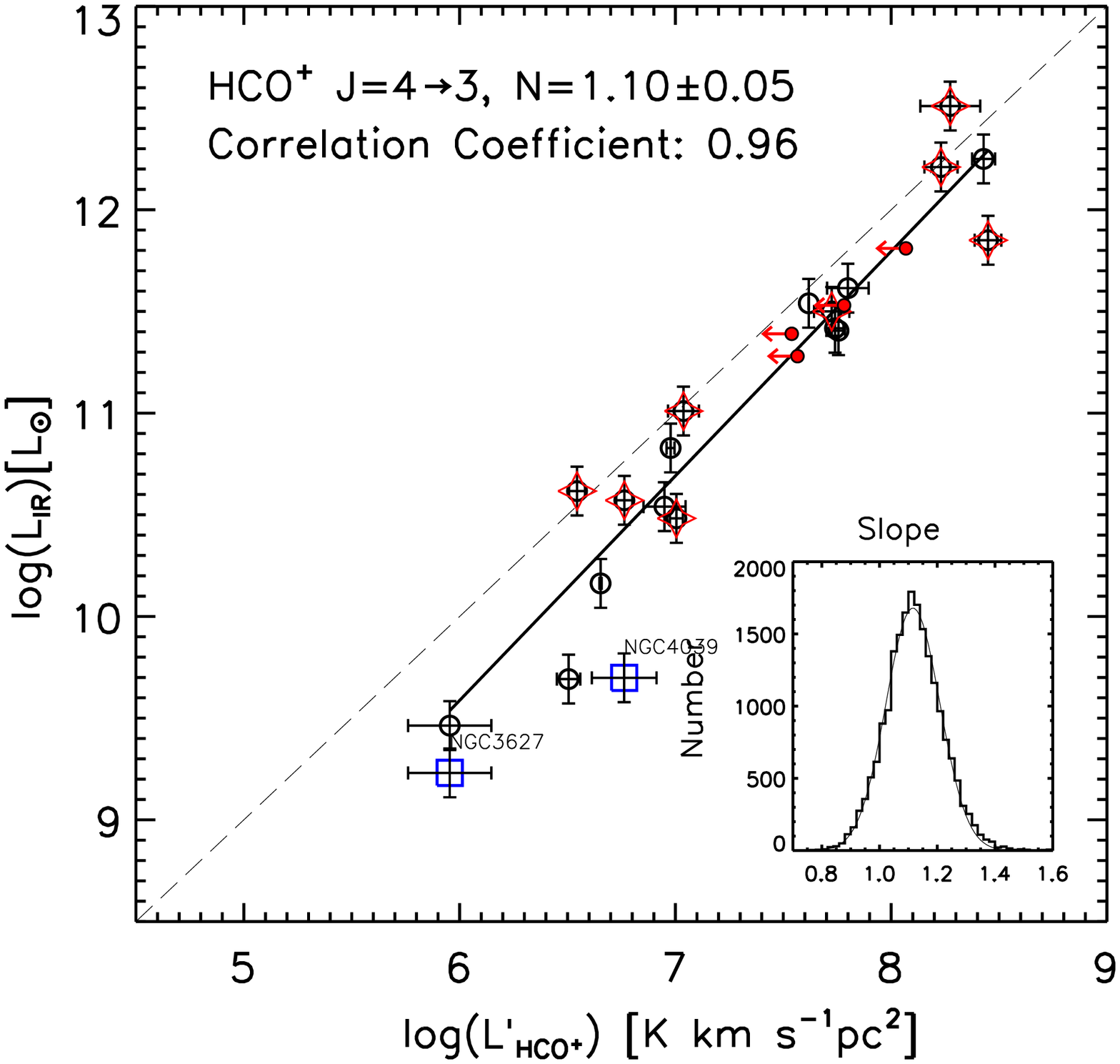}  
\includegraphics[angle=90,scale=0.4]{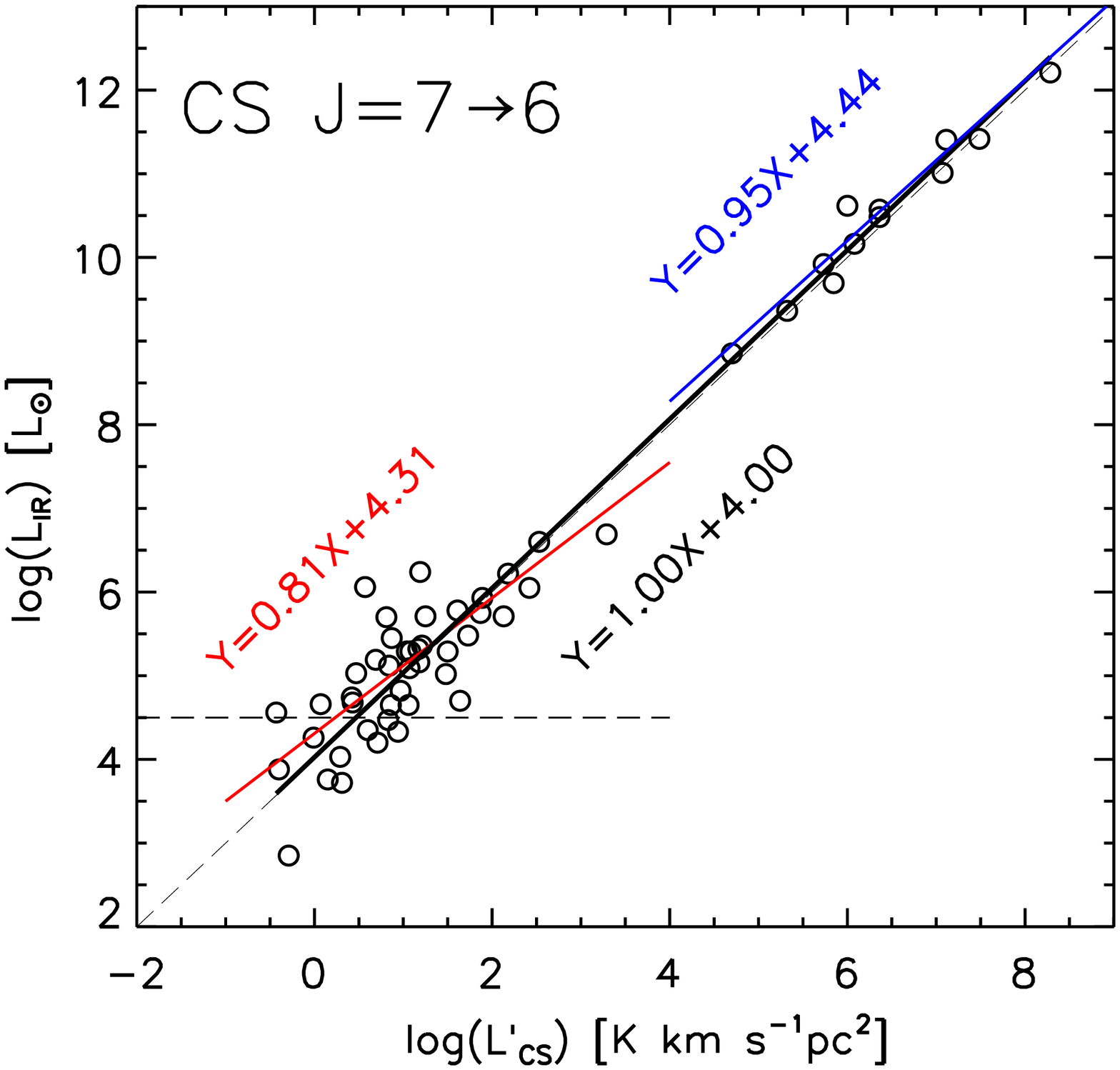}  
\caption{
Correlations between the gas luminosity log(\Lgas) and the IR luminosity
log(\LIR).  {\it Upper left:} \CSss.  {\it Upper right:} \HCNft.  {\it Bottom
left:} \HCOPft. {\it Bottom right:} \CSss\ for both galaxies and Galactic
cores. The horizontal dashed lines indicate \LIR\ = $10^{4.5}$ \Lsun, below
where the Galactic cores have less statistical meaning \citep[i.e.,][]{wes10}.
N is the slope of the correlation, and $r$ is the correlation
coefficient. The thick lines show the fitted results. The oblique dashed lines
indicate linear relations. Red empty diamonds are AGNs, open circles are
detections for the fitting, and filled circles and arrows are limits. NGC 4039
and NGC 3627 are not adopted for fitting and are plotted in blue. The insets
present probability density distributions of the slopes derived from the
Bayesian fitting.}
\label{correlations}
\end{figure*}


Our observations were performed in 2011 April and August with the Atacama
Pathfinder Experiment (APEX) on the Chajnantor Plateau in Chile, in good ($pwv
<$ 0.6 mm) to median ($pwv\sim$ 1 mm) weather conditions.  In total we
spent $\sim$ 30 hours telescope time on this project. The First Light APEX
Submillimeter Heterodyne receiver (FLASH) was employed to observe \CSss,
\HCNft\, and \HCOPft\ simultaneously, with dual sidebands.  Typical system
temperatures  were \Tsys$\sim$ 180 - 240\, K. The Fast Fourier
Transform Spectrometer back-ends led to a bandwidth of 4 GHz for each sideband,
with a channel spacing of 0.2\, MHz. The beam-size is $\sim18''$ at 342\, GHz. 

All observations were performed in a wobbler switching mode. Beam throws range
between 2$'$ and 4$'$, according to the target size. Every 15 minutes we made a
chopper wheel calibration. The focus was determined on Saturn or Jupiter every
3--6 hours. Pointing was checked once per hour, resulting in a typical
uncertainty of 2-3$\arcsec$ (R.M.S.). Including overhead, we spent $\sim$1.5
hour on each galaxy. Typical R.M.S. noise levels are 0.1 mK at 20 \kms\
velocity resolution. Although the sideband separation is better than 10
dB, there are still some CO $J$=3$\rightarrow$2 images presented in the upper
sideband. Fortunately these CO images do not mix with our HCN spectra.  

All data were reduced with the CLASS package in
GILDAS\footnote{\url{http://www.iram.fr/IRAMFR/GILDAS}}.  We checked the line
profiles of low-$J$ HCN or CO transitions in the literature and set the
baseline ranges accordingly. Linear baselines were subtracted after inspecting
each spectrum. We qualified spectra by comparing the measured noise and
the theoretical noise before and after 4 times of boxcar smooth.  About 5\% of
the spectra were discarded during the qualification.

We converted the antenna temperatures (\Tastar) to the main beam brightness
temperatures (\Tmb), using ${T_{\rm mb} = T_{\rm A}^{\star}\cdot F_{\rm
eff}/B_{\rm eff}}$. The adopted forward hemisphere efficiency $F_{\rm eff}$ and
beam efficiency $B_{\rm eff}$ are 0.95 and 0.73, respectively. The flux
uncertainty was estimated to be $\sim$15\%.  We derived the line luminosity
(\Lgas) following \citet[][]{Solomon1992Natur}. 

\section{RESULTS AND DISCUSSION}

We detected 8, 14, and 16 ($I_{\rm line}$$\ge$\ 3 $\sigma$) galaxies in \CSss,
\HCNft, and \HCOPft, respectively. Most of them are strong detections of HCN
and \HCOP\ ($\ge$\ 4 $\sigma$), while Mrk~331, NGC~7469 and NGC~4039 show
marginal detections in \HCOPft\ ($\sim$ 3 $\sigma$). Upper limits (3 $\sigma$)
are derived for the non-detections. Fig. \ref{spectra1} shows the spectra of
strong detections of \HCOPft, and Fig. \ref{spectra2} shows the spectra of weak
detections of \HCOPft. NGC~7771 and NGC~2903 has only non-detections, so we do
not show their spectra.  Combining data from the literature we construct
samples containing 14, 17, and 17 detected galaxies for \CSss, \HCNft, and
\HCOPft, respectively.

Total IR (TIR; 8-1000$\mu$m) luminosity \citep{smk03} is adopted as a
proxy of SFR.  However, the molecular lines are obtained only in small
beam-sizes ($\sim$ 14$''$ -- 18$''$), which only pick up a certain portion of
the IR luminosity for a whole galaxy.  This is particularly problematic for
nearby galaxies and starbursts because of their large radii (up to a few
arcmins). If this effect was neglected, one would systematically overestimate
the IR emission or underestimate the dense gas emission at the low luminosity
end of the \Lgas-\LIR\ correlations, while the ULIRGs are not affected due to
their small angular sizes of gas emission. 

We decide to measure the IR emission within the submm beam-size, rather
than to adopt the IR luminosities of the entire galaxies. We download {\it
Herschel}\footnote{Herschel is an ESA space observatory with science
instruments provided by European-led Principal Investigator consortia and with
important participation from NASA.} PACS 100 $\mu$m, 70 $\mu$m (when 100 $\mu$m
is not available) images from the Herschel Science Archive (HSA), and they are
processed to level 2.5 in the pipeline. NGC 3628 does not have {\it Herschel}
data, so we adopt archival images of SCUBA 850 $\mu$m instead. Using these IR
data, we perform aperture photometry with the submm beam-size and with the
whole galaxy. The background radii are selected by eye from the outside of
galaxies to the edge of the images. Some nearby galaxies are downloaded from
the Key Insights on Nearby Galaxies (KINGFISH) project, and similar to their
photometry results \citep{Dale2012}, the impact of using $\sim$10\% larger or
smaller aperture areas is a median difference of less than 3\% on the flux
densities for all wavelengths. The IR luminosity within the beam-size is: 

\[L_{\rm IR} = L_{\rm TIR} \times R_{\rm SD} \times C_{\rm aper} ,\] 

where $L_{\rm TIR}$ is the IR luminosity of the entire galaxy, $R_{\rm SD}$ is
the ratio of flux densities within the beam-size of single dish (SD)
telescope to that measured in the whole galaxy, and $C_{\rm aper}$ is the
aperture correction factor for the beam-sizes. The final error comprises errors
of photometry ($\sim$5\%), the point source assumption ($\sim$10\%), the flux
calibration error ($\sim$5\%), and the error of tracing TIR with a chromatic IR
band ($\sim$10\%) \citep[e.g.,][]{Galametz2013}. In the end we account 20\% as
a conservative uncertainty for \LIR. 

\subsection{The \Lgas-\LIR\ Correlations}

Figure \ref{correlations} presents the \Lgas-\LIR\, correlations. All
detections are included in the fitting except for NGC~4039 and NGC~3627,
because the regions covered by their APEX beams are not active in star
formation as indicated from their IR images. In the linear regression we assume
Gaussian independent variables, and account for the errors in both \LIR\, and
\Lgas.  Upper limits are not adopted in the fitting. The \Lgas-\LIR\,
correlations extend from the nuclear regions of nearby normal galaxies to
ULIRGs, covering a \LIR\ range of $\sim$ 2.5 decades. We adopt publicly
available IDL routines MPFIT \citep{Markwardt2009} for the linear least-squares
fit and LINMIX\_ERR \citep{Kelly2007} for the Bayesian regression.  The best
linear least-squares fits with uncertainties are listed below, with $r$
indicating the correlation coefficient:\\

\noindent
\[{\rm log}(L_{\rm IR})=0.95(\pm 0.04) \times {\rm log}(L'_{\rm CS   76})+4.33(\pm 0.28);r=0.99\] 
\[{\rm log}(L_{\rm IR})=1.00(\pm 0.04) \times {\rm log}(L'_{\rm HCN  43})+3.67(\pm 0.28);r=0.97\] 
\[{\rm log}(L_{\rm IR})=1.10(\pm 0.05) \times {\rm log}(L'_{\rm HCO^+43})+2.48(\pm 0.38);r=0.96\]

In the Bayesian regressions, we present the posterior distributions of
possible slopes in the insets of Figure \ref{correlations}, and obtain error
estimates by fitting the distributions with Gaussian profiles. These fittings
give slopes of 0.94$\pm$0.07, 1.01$\pm$0.07, and 1.12$\pm$0.9 for \CSss,
\HCNft, and \HCOPft, respectively. These are consistent to the linear
least-squares fit. Both correlations of \LCSss-\LIR\ and \LHCNft-\LIR\ have
slopes very close to unity, while the \LHCOPft-\LIR\ correlation shows a slight
super-linear slope. The slopes of HCN and \HCOP\ are much higher than
those predicted by \citet{ncs08} and \citet{jnm09}. Although they did not
predict \CSss, the high \ncrit\ of \CSss\ makes it far from the predicted trend
too. We also fitted the correlations with our APEX data only, and
obtained slopes of 0.98$\pm$0.15, 0.98$\pm$0.05, and 1.08$\pm$0.05, which are
very close to the results fitted with the combined data. 

Alternatively, we fitted the correlations without the beam-matching correction,
and obtained slopes (N) and correlation coefficients ($r$) to be:
N=0.71$\pm$0.08 and $r$=0.89 for \CSss, N=0.77$\pm$0.05 and $r$=0.94 for
\HCNft, and N=0.94$\pm$0.10 and $r$=0.92 for \HCOPft, respectively. The
shallower slopes (sub-linear) and worse correlations indicate that the
beam-matching correction has a significant effect in the fitting. On the other
hand it should be emphasized that, with the beam-matching correction we are
studying the central 14$''$-18$''$ regions in the nearby galaxies (except for
the maps of \HCNft\ and \HCOPft\ in NGC 253), rather than the whole galaxies.

At \Tkin=100K, \HCNft, \HCOPft, and \CSss\ have \ncrit\ of $5.6\times
10^6$\cmt, $1.3\times 10^6$\cmt, and $2.8\times10^6$ \cmt (see Sect
\ref{intro}), respectively. These tracers pick up the densest part of the
molecular cores, and trace only a small amount of the total dense gas mass
involved in star-formation. The linear slopes found in the high-$J$ HCN and CS
tend to support the proposed correlations for gas tracers with \ncrit\ higher
than $\sim 10^4$ \cmt\ \citep{Lada12}.   

Comparing with the neutral HCN and CS molecules, the molecular ion \HCOP\
is more sensitive to the average ambient electron abundance $x(e)$, because
of the protonation reaction: \[\rm H_3^+ + CO \rightarrow HCO^+ + H_2\]  The
high electron abundance is likely to reduce $\rm [HCO^{+}/H_2]$
efficiently. Considering the impacts of the CR ionization \citep{PPP2007}, the
turbulent diffusion \citep{Xie95}, and the AGN illumination, \HCOP\ is likely
deficit when being exposed to the extreme physical conditions prevailing at the
high luminosity end, i.e., ULIRGs. Such effect could increase the slopes of the
\LHCOP-\LIR  correlations in galaxies.

\subsection{Comparison between Galactic Cores and Galaxies}

\citet[][]{wes10} measured the \LCSss-\LIR\ correlation in individual Galactic
dense cores with high mass star formation. Their best fit yields a slope of
$\sim$ 0.8 with a large uncertainty because of the significant scatter in the
data and the limited range of infrared luminosities they have at their
disposal. In the bottom right of Figure \ref{correlations}, we plot their
Galactic cores and our galaxies together. Combining both samples, we find a
highly linear correlation, \[ {\rm log}(L_{\rm IR}) =1.00(\pm 0.01)\times {\rm
log}( L'_{\rm CS\, J=7-6})+4.03(\pm0.04),\] with a correlation coefficient of
0.98. This correlation is remarkably similar to that derived for HCN
$J=1\rightarrow$0 by \citet{Wu05}, though \CSss\ has much higher excitation
requirements. The linear correlation between \Lgas\ and \LIR\ holds over an IR
luminosity range of about eight orders of magnitude.

The linearity of the (dense gas)---SFR correlation was interpreted as a
``fundamental unit'' of star formation on different physical scales, thus both
SFR and dense gas mass are simply piled up by adding in more units
\citep[e.g.,][]{Wu05}. Recently it was found that once the studied gas content
is denser than a density threshold, SFR does not depend on the exact value of
the gas density, but depends on the total mass of the dense gas
\citep[e.g.,][]{Lada12}. In such a context the near-linearity in the
$\Sigma_{\rm gas}-\Sigma_{\rm SFR}$ correlation in nearby galaxies
\citep[e.g.,][]{Bigiel2008,Schruba12} is likely caused by a constant $M_{\rm
dense}$/$M_{\rm total\, H_2 }$ fraction. 

The gas densities probed by the three lines are beyond the highest average
densities in starbursts \citep{kt07}, and they are also one to two orders of
magnitude higher than the threshold density for star-formation \citep[\nht
$\sim 10^4$ \cmt; e.g., ][]{Parmentier11,Lada12}.  We find that all
gases with densities $> 10^4$ \cmt\ are linearly correlated with SFR, which
indicates that the SFR in the dense gas is not likely affected by the free-fall
time scale ($t_{\rm ff}$), because $t_{\rm ff} \propto \rho^{-1/2}$. If $\Sigma
_{\rm gas} / t_{\rm ff}$-$\Sigma _{\rm SFR}$ correlations are linear for all
dense gases, the shorter $t_{\rm ff}$ for the denser gas would not keep
\Lgas-\LIR\ linear.  Indeed, the gas content traced by \HCNoz\ has 10 times
longer $t_{\rm ff}$ than that traced by \HCNft, but they do show both linear
correlations with the SFR.  The non-linear correlations found in the K-S law
are likely caused by the different fractions of dense gas to total gas in
different types of galaxies, rather than the free-fall timescales
\citep{Lada12}.  Actually, the K-S law with a slope of $\sim$1.4 is likely not
a fundamental physical relation, as the non-linear slope found for H$_2$+\HI,
seen in the very tenuous gas, is related to the relative amounts of dense gas
and dust directly being involved in star formation and other gas being too
diffuse and diluted to have anything to do with massive star formation.

\subsection{AGN Contamination to the IR Emission} 

About half of the galaxies in our sample are hosting AGNs, in particular the
ULIRGs. Compared to the near- and mid IR bands, the far- and total-IR are well
correlated and are thus less contaminated by the AGNs emission
\citep[e.g.,][]{U2012,jnm09}. Some AGNs, however, can still contribute
significantly to the bolometric luminosity in ULIRGs
\citep[e.g.,][]{Veilleux2009}.  For the most extreme case, Mrk~231, the
estimated AGN contribution is $\sim$70\%, and the SFR indicated from the IR
emission should be $\sim$0.5 dex lower in the \Lgas-\LIR\ correlations
\citep[e.g.,][]{Veilleux2009}. For the rest galaxies, the AGN contamination
should be much less ($<$10\%) for LIRGs (\LIR $>10^{11}$\Lsun) and nearby
Seyferts, where star-formation is expected to dominate over IR emission. 
Unfortunately the IR emission for most AGN-host galaxies have not been
decomposed, so that we could not deduce the IR luminosity by $\eta_{\rm SF}$,
where $\eta_{\rm SF}$ is the fraction of the IR luminosity due to
star-formation. Overall, we do not find significant changes in the correlations
if we remove AGNs from our samples. This may be due to the small size of our
sample, or the fact that most AGN-hosting galaxies are not strongly
contaminated by AGNs.

In conclusion, we present a survey of three dense gas tracers (\CSss, \HCNft,
and \HCOPft) in 20  nearby galaxies observed with the APEX 12-m telescope.
Combining with data from literature and after a beam-matching correction for
nearby galaxies, we find linear \Lgas-\LIR\ correlations for \CSss\ and \HCNft,
and a slightly super-linear slope for \LHCOPft. These results are consistent
with those found in HCN\ $J$=1$\rightarrow$0 \citep[i.e.,][]{gs04b,gs04a} and
\CSff\ \citep[i.e.,][]{Wang2011,wes10}, but contradictory to the predictions in
\citet{bns08} and \citet{jnm09}. We also find that the linear
\LCSss-\LIR\ correlation can be traced $universally$ across eight orders of
luminosity magnitude, down to Galactic cores. If the beam-matching corrections
are not applied for nearby galaxies, however, the slopes are significantly
sub-linear. Future ALMA surveys of more dense gas tracers as well as other
transitions over large samples of galaxies are necessary to consolidate the
above findings and ensure that, apart from the vagaries of HCO$^{+}$ chemistry,
the simplest (dense gas)-SFR empirical relation is indeed true.

\begin{acknowledgements}
We are grateful to the MPfIR team and the APEX staff for their help and support
during and after the observations. We also thank the anonymous referee for
his/her very useful comments and suggestions which improved the quality of the
manuscript. ZZ thanks Padelis Papadopoulos, Thomas Greve, and Daizhong Liu for
their constructive suggestions. This work is supported by the NSFC grant
\#11173059.  This research has made use of the NASA/IPAC Extragalactic Database
(NED) which is operated by the Jet Propulsion Laboratory, California Institute
of Technology, under contract with the National Aeronautics and Space
Administration. \end{acknowledgements}

\bibliographystyle{apj}

\begin{thebibliography}{}


 \bibitem[{{Baan} {et~al.}(2008){Baan}, {Henkel}, {Loenen}, {Baudry}, \&
         {Wiklind}}]{bhl08} {Baan}, W.~A., {Henkel}, C., {Loenen}, A.~F.,
         {Baudry}, A., \& {Wiklind}, T.  2008, \aap, 477, 747
%
 \bibitem[{{Bayet} {et~al.}(2009){Bayet}, {Aladro},
         {Mart{\'{\i}}n}, {Viti}, \& {Mart{\'{\i}}n-Pintado}}]{bayet09}
         {Bayet}, E., {Aladro}, R., {Mart{\'{\i}}n}, S., {Viti}, S., \&
         {Mart{\'{\i}}n-Pintado}, J. 2009{\natexlab{a}}, \apj, 707, 126
%

 \bibitem[{{Bigiel} {et~al.}(2008){Bigiel}, {Leroy}, {Walter}, {Brinks}, {de
         Blok}, {Madore}, \& {Thornley}}]{Bigiel2008} {Bigiel}, F., {Leroy},
         A., {Walter}, F., {et~al.} 2008, \aj, 136, 2846
%
 \bibitem[{{Bussmann} {et~al.}(2008){Bussmann}, {Narayanan}, {Shirley},
         {Juneau}, {Wu}, {Solomon}, {Vanden Bout}, {Moustakas}, \&
         {Walker}}]{bns08} {Bussmann}, R.~S., {Narayanan}, D., {Shirley},
         Y.~L., {et~al.} 2008, \apjl, 681, L73
%
\bibitem[{{Dale} {et~al.}(2012){Dale}, {Aniano}, {Engelbracht}, {Hinz},
  {Krause}, {Montiel}, {Roussel}, {Appleton}, {Armus}, {Beir{\~a}o}, {Bolatto},
  {Brandl}, {Calzetti}, {Crocker}, {Croxall}, {Draine}, {Galametz}, {Gordon},
  {Groves}, {Hao}, {Helou}, {Hunt}, {Johnson}, {Kennicutt}, {Koda}, {Leroy},
  {Li}, {Meidt}, {Miller}, {Murphy}, {Rahman}, {Rix}, {Sandstrom}, {Sauvage},
  {Schinnerer}, {Skibba}, {Smith}, {Tabatabaei}, {Walter}, {Wilson}, {Wolfire},
  \& {Zibetti}}]{Dale2012}
{Dale}, D.~A., {Aniano}, G., {Engelbracht}, C.~W., {et~al.} 2012, \apj, 745, 95
%
 \bibitem[{{Evans}(2008)}]{e08} {Evans}, II, N.~J. 2008, in Astronomical
         Society of the Pacific Conference Series, Vol. 390, Pathways Through
         an Eclectic Universe, ed. J.~H. {Knapen}, T.~J. {Mahoney}, \&
         A.~{Vazdekis}, 52
%
\bibitem[{{Galametz} {et~al.}(2013){Galametz}, {Kennicutt}, {Calzetti},
        {Aniano}, {Draine}, {Boquien}, {Brandl}, {Croxall}, {Dale},
        {Engelbracht}, {Gordon}, {Groves}, {Hao}, {Helou}, {Hinz}, {Hunt}, {Johnson},
        {Li}, {Murphy}, {Roussel}, {Sandstrom}, {Skibba}, \&
        {Tabatabaei}}]{Galametz2013} {Galametz}, M., {Kennicutt}, R.~C., {Calzetti},
        D., {et~al.} 2013, \mnras,431,1956
%
 \bibitem[{{Gao} \& {Solomon}(2004{\natexlab{a}})}]{gs04b} {Gao}, Y. \&
         {Solomon}, P.~M. 2004{\natexlab{a}}, \apjs, 152, 63
%
 \bibitem[{{Gao} \& {Solomon}(2004{\natexlab{b}})}]{gs04a} {Gao}, Y. \&
         {Solomon}, P.~M. 2004{\natexlab{b}}, \apj, 606, 271
%
%
 \bibitem[{{Graci{\'a}-Carpio} {et~al.}(2008){Graci{\'a}-Carpio},
         {Garc{\'{\i}}a-Burillo}, {Planesas}, {Fuente}, \& {Usero}}]{gc08}
         {Graci{\'a}-Carpio}, J., {Garc{\'{\i}}a-Burillo}, S., {Planesas}, P.,
         {Fuente}, A., \& {Usero}, A. 2008, \aap, 479, 703
%
 \bibitem[{{Greve} {et~al.}(2009){Greve}, {Papadopoulos}, {Gao}, \&
         {Radford}}]{gpg09} {Greve}, T.~R., {Papadopoulos}, P.~P., {Gao}, Y.,
         \& {Radford}, S.~J.~E. 2009, \apj, 692, 1432
%
%
 \bibitem[{{Jackson} {et~al.}(1995){Jackson}, {Paglione}, {Carlstrom}, \&
         {Rieu}}]{Jackson95} {Jackson}, J.~M., {Paglione}, T.~A.~D.,
         {Carlstrom}, J.~E., \& {Rieu}, N.-Q.  1995, \apj, 438, 695
%
 \bibitem[{{Juneau} {et~al.}(2009){Juneau}, {Narayanan}, {Moustakas},
         {Shirley}, {Bussmann}, {Kennicutt}, \& {Vanden Bout}}]{jnm09}
         {Juneau}, S., {Narayanan}, D.~T., {Moustakas}, J., {et~al.} 2009,
         \apj, 707, 1217
%
 \bibitem[{{Kelly}(2007)}]{Kelly2007}
 {Kelly}, B.~C. 2007, \apj, 665, 1489
%
 \bibitem[{{Kennicutt}(1998)}]{Kennicutt98}
 {Kennicutt}, Jr., R.~C. 1998, \apj, 498, 541
%
 \bibitem[{{Knudsen} {et~al.}(2007){Knudsen}, {Walter}, {Weiss}, {Bolatto},
   {Riechers}, \& {Menten}}]{Knudsen07}
 {Knudsen}, K.~K., {Walter}, F., {Weiss}, A., {et~al.} 2007, \apj, 666, 156
%
 \bibitem[{{Krumholz} \& {Thompson}(2007)}]{kt07}
 {Krumholz}, M.~R. \& {Thompson}, T.~A. 2007, \apj, 669, 289
%
 \bibitem[{{Lada} {et~al.}(2012){Lada}, {Forbrich}, {Lombardi}, \&
   {Alves}}]{Lada12}
 {Lada}, C.~J., {Forbrich}, J., {Lombardi}, M., \& {Alves}, J.~F. 2012, \apj,
   745, 190
%
 \bibitem[{{Liu} \& {Gao}(2012)}]{Liu2012}
         {Liu}, L. \& {Gao}, Y. 2012, Science in China G: Physics and Astronomy, 55, 347
%
 \bibitem[{{Ma} {et~al.}(2012){Ma}, {Tan}, \& {Barnes}}]{Ma12}
         {Ma}, B., {Tan}, J.~C., \& {Barnes}, P.~J. 2012, ArXiv e-prints
%
\bibitem[{{Markwardt} (2009)
        {Markwardt}}]{Markwardt2009} {Markwardt}, C.~B.,  2009, ASPC, 411, 251 
%
%
 \bibitem[{{Narayanan} {et~al.}(2008){Narayanan}, {Cox}, {Shirley}, {Dav{\'e}},
   {Hernquist}, \& {Walker}}]{ncs08}
 {Narayanan}, D., {Cox}, T.~J., {Shirley}, Y., {et~al.} 2008, \apj, 684, 996
%
 \bibitem[{{Papadopoulos}(2007)}]{PPP2007} {Papadopoulos}, P.~P. 2007, \apj,
         656, 792
%
 \bibitem[{{Parmentier} {et~al.}(2011){Parmentier}, {Kauffmann}, {Pillai}, \&
         {Menten}}]{Parmentier11} {Parmentier}, G., {Kauffmann}, J., {Pillai},
         T., \& {Menten}, K.~M. 2011, \mnras, 416, 783
%
 \bibitem[{{Reiter} {et~al.}(2011){Reiter}, {Shirley}, {Wu}, {Brogan},
         {Wootten}, \& {Tatematsu}}]{Reiter2011} {Reiter}, M., {Shirley},
         Y.~L., {Wu}, J., {et~al.} 2011, \apjs, 195, 1
%
%
 \bibitem[{{Sanders} {et~al.}(2003){Sanders}, {Mazzarella}, {Kim}, {Surace}, \&
         {Soifer}}]{smk03} {Sanders}, D.~B., {Mazzarella}, J.~M., {Kim}, D.-C.,
         {Surace}, J.~A., \& {Soifer}, B.~T. 2003, \aj, 126, 1607
%
 \bibitem[{{Schenck} {et~al.}(2011){Schenck}, {Shirley}, {Reiter}, \&
         {Juneau}}]{Schenck11} {Schenck}, D.~E., {Shirley}, Y.~L., {Reiter},
         M., \& {Juneau}, S. 2011, \aj, 142, 94
%
 \bibitem[{{Sch{\"o}ier} {et~al.}(2005){Sch{\"o}ier}, {van der Tak}, {van
         Dishoeck}, \& {Black}}]{schoier2005} {Sch{\"o}ier}, F.~L., {van der
         Tak}, F.~F.~S., {van Dishoeck}, E.~F., \& {Black}, J.~H. 2005, \aap,
         432, 369
%
 \bibitem[{{Schruba} {et~al.}(2012){Schruba}, {Leroy}, {Walter}, {Bigiel},
         {Brinks}, {de Blok}, {Kramer}, {Rosolowsky}, {Sandstrom}, {Schuster},
         {Usero}, {Weiss}, \& {Wiesemeyer}}]{Schruba12} {Schruba}, A., {Leroy},
         A.~K., {Walter}, F., {et~al.} 2012, \aj, 143, 138
%
 \bibitem[{{Solomon} {et~al.}(1992){Solomon}, {Radford}, \&
         {Downes}}]{Solomon1992Natur} {Solomon}, P.~M., {Radford}, S.~J.~E., \&
         {Downes}, D. 1992, \nat, 356, 318
%
 \bibitem[{{Spergel} {et~al.}(2007){Spergel}, {Bean}, {Dor{\'e}}, {Nolta},
         {Bennett}, {Dunkley}, {Hinshaw}, {Jarosik}, {Komatsu}, {Page},
         {Peiris}, {Verde}, {Halpern}, {Hill}, {Kogut}, {Limon}, {Meyer},
         {Odegard}, {Tucker}, {Weiland}, {Wollack}, \& {Wright}}]{sbd07}
         {Spergel}, D.~N., {Bean}, R., {Dor{\'e}}, O., {et~al.} 2007, \apjs,
         170, 377
%
 \bibitem[{{U} {et~al.}(2012){U}, {Sanders}, {Mazzarella}, {Evans}, {Howell},
         {Surace}, {Armus}, {Iwasawa}, {Kim}, {Casey}, {Vavilkin}, {Dufault},
         {Larson}, {Barnes}, {Chan}, {Frayer}, {Haan}, {Inami}, {Ishida},
         {Kartaltepe}, {Melbourne}, \& {Petric}}]{U2012} {U}, V., {Sanders},
         D.~B., {Mazzarella}, J.~M., {et~al.} 2012, \apjs, 203, 9
%
 \bibitem[{{Veilleux} {et~al.}(2009) {Veilleux}, {Rupke}, {Kim}, {Genzel},
         {Sturm}, {Lutz}, {Contursi}, {Schweitzer}, {Tacconi}, {Netzer},
         {Sternberg}, {Mihos}, {Baker}, {Mazzarella}, {Lord}, {Sanders},
         {Stockton}, {Joseph}, \& {Barnes}}]{Veilleux2009} {Veilleux}, S.,
         {Rupke}, D. S. N., {Kim}, D.-C., {Genzel}, R., {et~al.} 2009, \apjs,
         182, 628 
%
 \bibitem[{{Wang} {et~al.}(2011){Wang}, {Zhang}, \& {Shi}}]{Wang2011} {Wang},
         J., {Zhang}, Z., \& {Shi}, Y. 2011, \mnras, 416, L21
%
 \bibitem[{{Wilson} {et~al.}(2008){Wilson}, {Petitpas}, {Iono}, {Baker},
         {Peck}, {Krips}, {Warren}, {Golding}, {Atkinson}, {Armus}, {Cox},
         {Ho}, {Juvela}, {Matsushita}, {Mihos}, {Pihlstrom}, \&
         {Yun}}]{Wilson08} {Wilson}, C.~D., {Petitpas}, G.~R., {Iono}, D.,
         {et~al.} 2008, \apjs, 178, 189
%
 \bibitem[{{Wu} {et~al.}(2005){Wu}, {Evans}, {Gao}, {Solomon}, {Shirley}, \&
         {Vanden Bout}}]{Wu05} {Wu}, J., {Evans}, II, N.~J., {Gao}, Y.,
         {et~al.} 2005, \apjl, 635, L173
%
 \bibitem[{{Wu} {et~al.}(2010){Wu}, {Evans}, {Shirley}, \& {Knez}}]{wes10}
         {Wu}, J., {Evans}, II, N.~J., {Shirley}, Y.~L., \& {Knez}, C. 2010,
         \apjs, 188, 313
%
 \bibitem[{{Xie} {et~al.}(1995){Xie}, {Allen}, \& {Langer}}]{Xie95} {Xie}, T.,
         {Allen}, M., \& {Langer}, W.~D. 1995, \apj, 440, 674

\end{thebibliography}

\end{document}